# Interactive Ant Colony Optimization (iACO) for Early Lifecycle Software Design

Christopher L. Simons, Jim Smith and Paul White

*Department of Computer Science and Creative Technologies*
*University of the West of England*
*Bristol  BS16 1QY  United Kingdom*
{chris.simons, james.smith, paul.white}@uwe.ac.uk

**Abstract**

Software design is crucial to successful software development, yet is a demanding multi-objective problem for software engineers. In an attempt to assist the software designer, interactive (i.e. human in-the-loop) meta-heuristic search techniques such as evolutionary computing have been applied and show promising results. Recent investigations have also shown that Ant Colony Optimization (ACO) can outperform evolutionary computing as a potential search engine for interactive software design. With a limited computational budget, ACO produces superior candidate design solutions in a smaller number of iterations. Building on these findings, we propose a novel interactive ACO (iACO) approach to assist the designer in early lifecycle software design, in which the search is steered jointly by subjective designer evaluation as well as machine fitness functions relating the structural integrity and surrogate elegance of software designs. Results show that iACO is speedy, responsive and highly effective in enabling interactive, dynamic multi-objective search in early lifecycle software design. Study participants rate the iACO search experience as compelling. Results of machine learning of fitness measure weightings indicate that software design elegance does indeed play a significant role in designer evaluation of candidate software design. We conclude that the evenness of the number of attributes and methods among classes (NAC) is a significant surrogate elegance measure, which in turn suggests that this evenness of distribution, when combined with structural integrity, is an implicit but crucial component of effective early lifecycle software design.

# 1    INTRODUCTION

Software design is both fundamental to the successful development of software-intensive systems and cognitively demanding for software engineers to perform. Indeed, in early-lifecycle software design, designers wrestle with numerous trade-off judgments as they formulate candidate design solutions as a basis for subsequent down-stream development activities. In an attempt to assist the software designer, interactive meta-heuristic search techniques such as evolutionary algorithms (EAs) with the software designer 'in-the-loop' have been recently studied and show promising results. After early empirical investigations incorporating designer preferences in search [1], [2], subsequent studies have combined quantitative machine-calculated fitness functions with qualitative designer evaluation of design elegance in a dynamic, multi-objective, interactive search [3]. These studies show that the precise balance of factors affecting the subjective judgments of the human software designer is highly significant but poorly understood – hence the oft-heard references to the "art" of software design.

Interestingly, however, recent investigations comparing different meta-heuristic search approaches have shown that ant colony optimization (ACO) can outperform evolutionary computation in quantitative search with respect to arriving at design solution candidates of superior fitness at earlier iterations / generations [4], [5]. This suggests that as an engine for interactive search, ACO has great potential. One major contribution of this paper is to exploit this potential by surveying a range of approaches to interactive multi-objective search, and then making an informed proposal for interactive ACO (iACO) for software design (Section 3). To evaluate the proposal, we describe the experimental methodology for an empirical study involving a number of software engineers in three case studies of early lifecycle software design (Section 4). Results of empirical investigations are presented in Section 5, while threats to validity are discussed in Section 6. Finally in Section 7, we conclude by assessing the effectiveness of iACO in assisting the software designer in early lifecycle software design.

# 2 BACKGROUND

## 2.1 Search-Based Early-Cycle Software Design

From its early roots using genetic algorithms to evolve software test sequences [6], [7], the idea that many aspects of software development are essentially optimization problems, and as such are amenable to automated search, has rapidly gained currency. In most cases the search suffers from combinatorial explosion, and the "fitness" landscapes are thought to exhibit discontinuities and multiple optima, motivating the use of meta-heuristics to perform the search. The term "Search Based Software Engineering" (SBSE) was coined around the turn of the millennium by Harman and Jones [8]. In the last decade applications of SBSE can be found across the spectrum of the software development lifecycle, including requirements analysis and scheduling [9], design tools and techniques [2], [10], software testing [11], automated bug fixing [12], and software maintenance [13]. A comprehensive repository of publications in SBSE is maintained by Zhang [14].

In early-lifecycle software development, it is necessary to first define requirements for the software system-to-be relevant to the problem domain under investigation. Then the designer identifies and evaluates concepts and information relevant to the design problem domain. This is an intensely people-centric activity, and typically involves multi-objective trade-offs using competing criteria [15], [16], [17]. Clearly, such design trade-offs are largely subjective, depending greatly on the competence of the individuals performing the design. Using the object-oriented paradigm, the identified concepts and information are expressed as "objects" and "classes" and these constructs have crucial relevance to subsequent downstream software implementation and testing. The Unified Modelling Language (UML) [18] is widely used by software designers to visualize and specify classes as well as other aspects of software designs. Using the UML, classes are placeholders or groupings of attributes (i.e. data that need to be stored, computed and accessed), and methods (i.e. units of execution by which objects communicate with other objects, programs and users). Thus early lifecycle software design can be formulated as a search among possible design structures for those comprising an appropriate grouping of attributes and methods into classes.

## 2.2 Interactive Meta-Heuristic Optimization

Fundamentally, the aim of interactive meta-heuristic search in early lifecycle software design is to support rather than replace the designer. Indeed, interactive EAs have been successfully applied in a wide range of applications to facilitate user-personalization. Typically the user is presented with a number of solutions, and rates them according to how well they meet their desiderata. This process implicitly captures the user's multi-objective decision making processes without the need for time consuming explicit knowledge-acquisition process [19]. Well known early applications include face-recognition [20], the evolution of computer graphics [21], and fitting Cochlear Implants [22].

Interactive multi-objective search techniques have also been widely used in the Multi-Criteria Decision Making (MCDM) community to gain insight into combinatorial optimization problems. Miettinen [23] provides a comprehensive survey of interactive search methods and distinguishes various methods of decision-maker involvement in multi-objective search, such as *a priori* methods: *"where the decision maker must specify their preferences, hopes and opinions"* before the automated search, as opposed to *a posteriori* methods, which perform automated search to proceed without human guidance, then provide the decision maker with a selection of alternative solutions. Both these methods are differentiated from interactive search wherein the human actively participates in the on-going search process. Belton *et al.* [24] examine interactive multi-objective optimization from a learning perspective, and speculate on ways to enable mutual learning between decision makers and search processes while emphasizing the role of interactive decision making software tools and environments. Deb attempts to consolidate knowledge of the MCDM and SBSE communities to assess the state-of-the-art in evolutionary multi-objective optimization [25]. Deb also considers multi-objective user evaluation in search and highlights the ***need for a dynamic search process in which objectives, constraints and search parameters may change over time to suit the interaction of the individual*** (our emphasis).

## 2.3 Reducing the Cognitive Burden of Interactive Search

The reliance on human guidance and judgment to direct and control the search, presents both potential weaknesses and strengths. On one hand, human subjective assessment tends to have a component of inconsistency and non-linearity of focus over time [26], which creates a need for rapid convergence. On the other hand, the ability to swiftly maneuver the search interactively can be exploited as a powerful strategy for adapting an otherwise naive search process. There have been a number of studies addressing the issues related to minimizing fatigue, both physical and psychological, that can result from prolonged interaction times and the possible stress of the evaluation process. Discretizing continuous fitness values into five or seven levels was shown to facilitate decision making, without significantly compromising convergence [27]. This limit on capacity for processing information has been comprehensively discussed in Miller [28] where he suggests organizing the information successively into a sequence of "chunks" to help stretch this limit on bandwidth.

An alternative approach to reducing time taken to discover good solutions is combining larger population sizes with a screening mechanism in which only a few individuals showing superior fitness are displayed to the user. Several methods have been proposed as "surrogate models" of user-provided fitness by, for example, clustering individuals [29], [30] or using multiple fuzzy state-value functions to approximate the trajectory of human scoring [31]. Avigad *et al.* [32], propose a multi-objective EA in which a model-based fitness of sub-concept solutions (using a sorting and ranking procedure) is combined with human evaluation. Similar approaches are reported by Brintrup *et al.* [33]. Previously [3] we have used periodic qualitative (user-provided) evaluations of software designs to dynamically update a surrogate model that combined quantitative measurements of structural integrity and metrics relating to design symmetry to drive the evolution of elegant software designs with reduced need for human evaluations.

Of course, a computationally efficient search engine is a prerequisite for a compelling interactive search experience. To minimize any frustration and/or fatigue for the user, and to maximize the consistency of user interaction, the underlying computational search must achieve a number of characteristics (see e.g. [19], [23]). Firstly, it must effectively explore the search space to arrive at candidate solutions of superior fitness, while at the same time allowing the search to be jointly steered by subjective user evaluation. Secondly, it must produce superior candidate design solutions in a minimum number of search iterations to provide a sense of positive progress for the user. Thirdly, it must be capable of multi-objective search, and be dynamically sensitive to user evaluation.

## 2.4 Choice of Meta-Heuristics for Interactive Search

Evolutionary computing is well understood and has a long history of success in interactive search, but history *per se* is not necessarily a good scientific motivation for investigation. With this in mind, the multi-objective performance of evolutionary algorithms (EAs, e.g. [34]) and ant colony optimization (Simple-ACO or S-ACO [35], [36]) have been compared by Simons and Smith for software design, with respect to both structural integrity and surrogate elegance metrics [4], [5]. The results are summarized as follows. Given a large computational budget (in terms of search iterations), an evolutionary algorithm with an integer-based representation emerges as superior. The evolutionary algorithm is also more robust for very large scale design problems where the number of classes in a software design is high. However, if the computational budget is limited (as is likely in interactive search), then a very different picture emerges. In this case, using a graph representation of software designs, ACO finds higher quality solutions, and in less search iterations. Moreover, in design solutions where the number of classes is fewer (but nevertheless typical of a realistic design problem) and the number of attributes and methods are high, ACO discovers candidate design solutions in approximately half the number of search iterations of the evolutionary algorithm. Simons and Smith conclude that ACO has significant potential as a search engine for interactive software design.

It is perhaps surprising to note that very few examples of interactive search involving ACO appear in the research literature. An early attempt to apply ACO to the design of constrained engineering design problems is reported in [37]. Some years pass before there is a report of interactive search with Particle Swarm Optimization used to design temperature profiles for a batch beer fermenter in 2005 [38]. Xing *et al.* [39] report the use of interactive fuzzy ACO for Job Shop Problems in 2007, while Ugur and Aydin describe an interactive simulation for solving the TSP using ACO in 2009 [40]. More recently, Albakour *et al.* report the use of ACO to simulate and interact with user query logs to learn knowledge about user behavior in a collection of documents [41].

Notwithstanding the above, it would seem that reports of interactive ACO used in any design domain are not abundant in the literature. Nevertheless, our previous encouraging results of ACO as a search engine strongly suggest that application of ACO for interactive software design shows great potential.

# 3  PROPOSED APPROACH

In this section, the software design problem and solution representations of the proposed approach are described, and this is followed by a specification of the fitness measures used. Next, the iACO search engine is described. Lastly, the proposed approach to software designer interaction is outlined to show how the designer's qualitative evaluations are integrated with the iACO search engine.

## 3.1  Representation

The software design problem is specified by means of UML use cases i.e. scenarios of usage of the software system-to-be described in terms of the interactions of humans (as actors) with the automated system [18]. The natural language text of the use case descriptions is analyzed as follows. Nouns are identified as data; verbs are identified as actions. If a datum is acted upon by the action, as is typical when the datum and action are co-located in a single interaction in the use case scenario, the action is said to "use" the datum. Thus in the language of UML, a software design problem is defined by a set of "methods" (actions), a set of "attributes" (data), and their corresponding "uses". This mapping ensures traceability from the design problem to the design solution. A full description of this software design problem specification can be found at [2].

The software design solution representation used is inspired by models for the Travelling Salesman Problem (TSP) and Vehicle Routing Problem (VRP) [42]. A solution consists of a complete path through a graph whose vertices represent elements of a software design solution. These are all of the attributes and methods, and we also add "*end of class*" elements (akin to "*return to depot*" markers) to delimit the scope of individual classes in the design solution path.

## 3.2  Fitness Measures

To reflect the multi-objective nature of the ACO search, a combination of fitness measures is used. The first fitness measure provides an assessment of the structural integrity of a software design. Designers typically strive for high cohesion in classes (to reflect a clear purpose) and low coupling between objects (to ensure the design is robust yet flexible to change). Thus, the first fitness measure is inspired by the "Coupling Between Objects" (CBO) measure [43]. For each candidate solution path, the CBO is calculated as the proportion of all "uses" of attributes by methods that occur across class boundaries. Thus, conveniently, a completely de-coupled design (all uses occur inside classes) scores a CBO of 1.0 while a completely coupled design scores a CBO of zero.

The second and third fitness measures provide an assessment of the elegance of the software design. We have previously proposed and investigated four novel quantitative elegance metrics relating to the evenness of distribution of attributes and methods among classes within the design [3]. That analysis revealed that two are particularly useful and effective, and it is these that we use as surrogates for human qualitative elegance evaluation. They are:

- *Numbers Among Classes (NAC)* is the arithmetic mean of the standard deviations of the numbers of attributes, and of methods among the classes of a design. The notion here is that the lower the value for NAC elegance, the more symmetrical the appearance of attributes and methods among the classes in the design as a whole.

- *Attribute to Method Ratio (ATMR)* is the standard deviation of the ratio of attributes to methods among the classes in a design. The notion here is that the lower the value of ATMR elegance, the more symmetrical the appearance of attributes and methods in individual classes of the design.

Good solutions are obtained through the minimization of CBO, NAC and ATMR.

### 3.3 iACO Search Engine

The design of the proposed iACO search engine has been influenced by the results of previous recent studies [44], [45], [46] and also draws inspiration from the *MAX-MIN* Ant System (*MM*AS) [47]. Indeed, three aspects of the elitist *MM*AS have been incorporated into the proposed iACO search engine. Firstly, only the iteration-best ant, i.e. the ant that produced the best candidate solution path in the current iteration, deposits pheromone. Secondly, the possible range of pheromone trail values are limited to an interval [$t_{min}$, $t_{max}$], and thirdly, pheromone trails are initialized to the upper trail limit i.e. $t_{max}$. However, the variant of *MM*AS used in this study does differ from the original *MM*AS in two respects to meet the needs of the software design domain. Firstly, local search is not conducted at each iteration and secondly, the influence of pheromone update is controlled by an additional parameter, $\mu$.

### 3.4 Software Designer Interaction

Providing effective interactive search for the software designer requires that we address the following questions:

*3.4.1 What implicit factors underlie the user's value judgments?*

The underlying value judgment made by the software designer relates to the trade-off between structural integrity in terms of class cohesion and design coupling, and design elegance. Having been presented with a visualization of a candidate software design solution, the software designer is invited to provide an overall evaluation on a scale of 1 to 100 where 1 is poor and 100 is good. With the twin aims of (i) reducing the number of interactions, and (ii) increasing our understanding of this value judgment process, the iACO uses a surrogate fitness model whose parameters are adapted in response to the periodic user evaluations. Historically interactive EAs have ranged between presenting a single individual for evaluation, to presenting the entire population for ranking [19], [23]. This continuum, especially ranking, clearly makes increasing cognitive demands of the user. Moreover, a single software design solution will typically be semantically rich in terms of design information, therefore since the iACO primarily uses the surrogate model for fitness, we present a single solution for evaluation selected at random from the set of non-dominated solutions within the population. The specific method adopted for the surrogate model is multiple linear regression:

$Predicted\_User\_Evaluation = a_0 + a_1 * CBO + a_2 * NAC + a_3 * ATMR$  (1)

where $a_0, a_1, a_2$ and $a_3$ are constants, initialized to 0, 0.34, 0.33, 0.33 respectively and thereafter updated whenever new observations become available. The new weights for each fitness function are then calculated in proportion to their coefficient i.e. a weighted sum approach.

*3.4.2 How are candidate solutions presented to the designer?*

Candidate solutions are presented in the form of UML class diagrams. Since color has been found to play an important role in design visualization [2], [3] it is used to reflect one aspect of the relative fitness of classes presented. It is proposed to trial the use of two color metaphors in this study: 'traffic lights' and 'water tap'. Classes with high, intermediate or low cohesion are coloured respectively in green/amber/red (traffic light) or red/amber/blue (water taps). Couples between classes are shown graphically as an unbroken line, with an arrowhead showing the direction of the couple. The stronger the coupling between classes, the thicker the line used. Examples of software design solution visualizations are available at [48].

*3.4.3 When does the designer interact with search?*

The crucial issue here is that user fatigue and loss of consistency places a limited "budget" on the number of interactions, which must be spent wisely. The starting point of the overall search process is the first iteration of *MM*AS wherein the generation of design solution paths is at random. However, previous work has shown that the multi-objective ACO search engine requires possibly 50 iterations to achieve reasonable fitness with respect to the three measures [5]. Moreover, using surrogate models makes it unnecessary for the designer to interact at each ACO iteration. Hence a better approach is to encourage a sense of positive progress in ACO search and enable designer interaction after an interval of several ACO iterations. Building on promising previous work [3], we employ an adaptive, fitness proportionate iteration interval. When poor values are observed for fitness measures, the scheme produces a high iteration interval (corresponding to 10 to 15 ACO iterations), as fitness measures improve, the iteration interval decreases. This allows the ACO search engine to speedily explore the search space, but also allows the designer evaluation to increasingly influence the direction of search as the interactive episode progresses.

Miettinen ([23] p. 134) provides three stopping criteria: *"Either the decision maker gets tired of the solution process, some algorithmic stopping (convergence) rule is fulfilled, or the decision maker finds a desirable solution and wants to stop. It is difficult to define precisely when a solution is desirable enough to become a final solution".* In this work, stopping is entirely at the discretion of the software designer.

*3.4.4 What means are provided to promote designer learning?*

Several mechanisms are provided to promote designer learning, and are centered on the notion of the designer having the opportunity to provide 'hints' to the iACO search engine. For example, it is possible for the designer to focus on individual classes of the design solution considered interesting and useful, and 'freeze' the classes with respect to on-going search. In terms of the evolving search, the designer is mentally *"anchoring"* i.e. fixing their thinking on some bias or partial 'chunk' of the solution [49]. It is also possible for the designer to 'unfreeze' class(es) at any interaction. This 'freezing' mechanism also provides an effective mechanism for the designer to address larger scale designs – smaller 'chunks' of the solution can be controlled before moving onto further design chunks. An additional designer learning mechanism is the ability to place interesting and useful software design solutions into an archive as iACO search progresses. This archive recall and comparison of interesting design solutions diminishes cognitive burden and promotes learning. A flow chart of the iACO algorithm is shown in Figure 1.

Figure 1. Flow chart of proposed dynamic multi-objective iACO Search. Sequential activities are shown in rectangles with solid lines; decisions and optional activities are shown with dotted lines

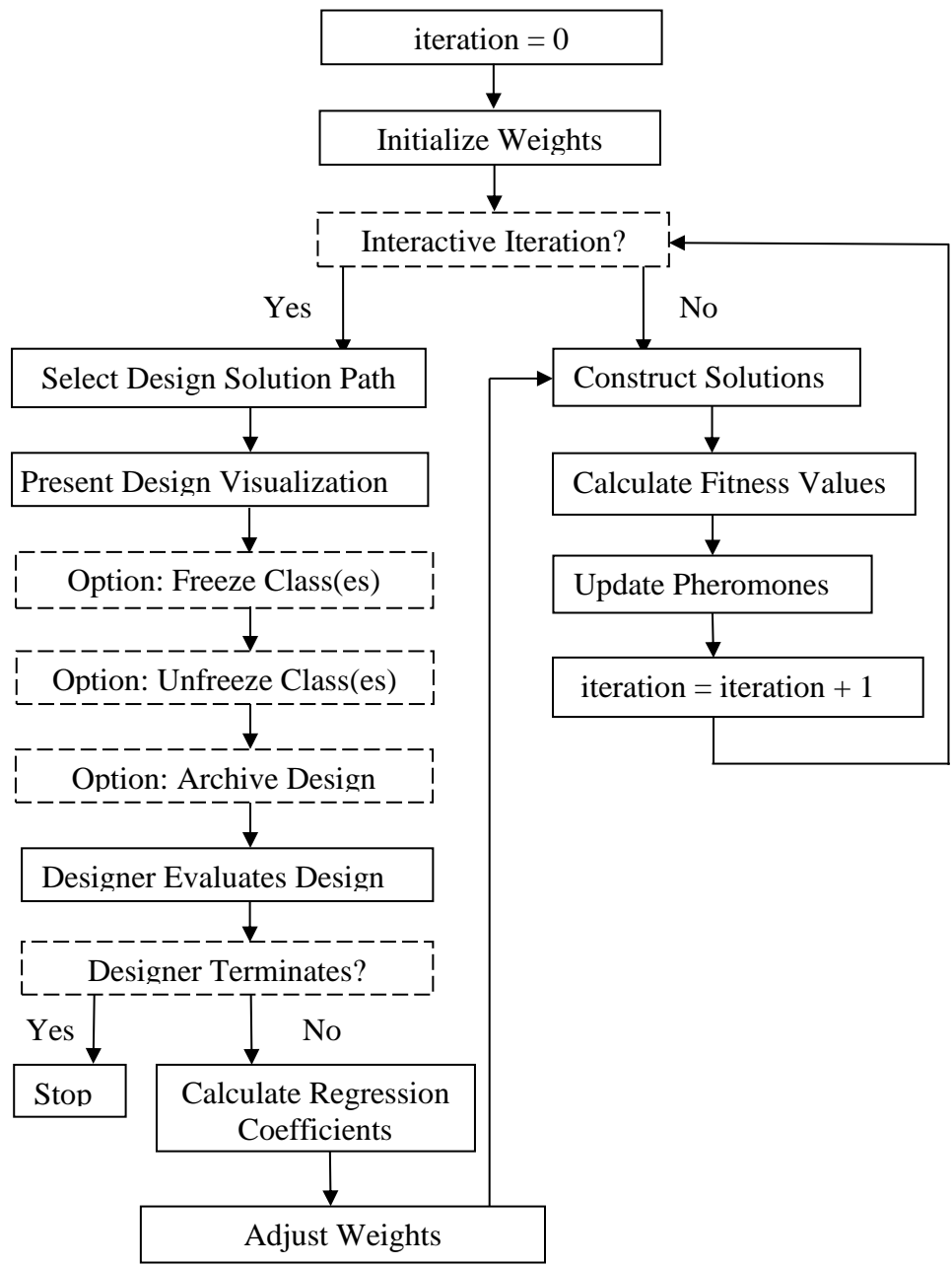

With respect to multi-objective ACO, a number of possibilities have been considered and evaluated for pheromone update, including weighted sum, weighted product pheromone update, multiple ant colonies and/or multiple pheromone matrices [44], [46]. However, given the requirement that the proposed iACO search engine be capable of dynamically adapting to the value judgments (evaluation) of the software designer during search while providing speedy search, computational straightforwardness and speed of execution is the priority and so the design choice of multiple ant colonies and multiple matrices has been rejected at this stage. In addition, empirical investigations to compare the performance of weighted sum versus weighted product pheromone update have been conducted for the three case studies used in the paper (see next section). For the sake of brevity, we can report that empirical results show that weighted sum update performance is marginally superior to weighted product update for the software design problem. (This finding is interesting insofar as it differs from the findings of Lopez-Ibanez and Stutzle on bi-objective TSP problem [46] with respect to approximation of the center of the Pareto front.)

## 4    EXPERIMENTAL METHODOLOGY

In this section, we first describe our choice of software design problem domains for use in our experiments. Then, secondly, we state algorithm parameters used. Lastly, we describe our methodology for an empirical investigation to assess the performance of iACO when used by software designers from a variety of backgrounds and experience.

### 4.1    Software Design Problems

A useful discussion on the choice of test problems for experimental comparison in meta-heuristic search is provided by Eiben and Smith ([50], pp.252-258), who compare and contrast the use of predefined problem instances (e.g. benchmark problems), problem instance generators, and 'real world' problems. Clearly, the use of predefined benchmark problems is preferable. Unfortunately, we are not aware of the existence of any recognized benchmark software design problems, either in the research literature or from industrial practice. It would be possible to generate instances of design problems, for example, with randomly defined attributes and methods. However, this presents the problem of semantics and understanding for the designer – it is likely that the generated design problem would be meaningless. Therefore, we have selected three real world software design problems from a variety of domains, with a range of scale.

      The first problem is a generalized abstraction of a Cinema Booking System (CBS), which addresses, for example, making an advance booking for a showing of a film at a cinema, and payment for tickets on attending the cinema auditorium. A specification of the use cases of is available at [51]. The second problem is an extension to a student administration system performed by the in-house information systems department at the authors' university. The University

sought to record outcomes relating to its Graduate Development Program (GDP). The extension was implemented and deployed in 2008. A specification of the use cases used in the development is available from [52]. The third problem is based on an industrial case study – Select Cruises (SC) - relating to a company selling nautical adventure holidays on tall-masted ships. The automated system handles quotation requests, cruise reservations, payment and confirmation via paper letter mailing. A specification of the use cases is available at [53]. Manual software designs have been performed by appropriate experienced software designers from the three industrial problem domains and are available from [54]. The manually performed designs for CBS and GDP show 5 classes while the manually performed design for SC shows 15 classes, and so the numbers of classes in the design solutions presented in the iACO environment is the same. Table 1 shows the number of attributes, methods and uses for each design problem and the values for different fitness metrics for the manual design.

Table 1. Scale of Software Design Problems and fitness value for manual design

| Problem | Attributes | Methods | Uses | CBO | NAC | ATMR |
|---------|------------|---------|------|-------|-------|-------|
| CBS | 16 | 15 | 39 | 0.154 | 0.821 | 0.199 |
| GDP | 43 | 12 | 121 | 0.297 | 2.592 | 2.617 |
| SC | 52 | 30 | 126 | 0.452 | 1.520 | 1.848 |

### 4.2   Algorithm Parameters

Values in Table 2 for the parameters $N$, $α$, $μ$, $σ$, are derived from the promising performances reported in [5], while those for $MM$AS-specific $t_{min}$ and $t_{max}$ are based on [47] and empirical evaluation with respect to the chosen representation.

Table 2. Algorithm Parameter Values.

| Parameter | Description | Value |
|-----------|-------------|-------|
| $N$ | Number of ants in colony | 100 |
| $α$ | Attractiveness of pheromone trails | 1.5 |
| $μ$ | Update of pheromone trails | 3.0 |
| $\sigma$ | Evaporation rate of pheromone trails | 0.035 |
| $t_{\min}$ | Minimum trail value within $MM$AS. | 0.5 |
| $t_{\max}$ | Maximum trail value within $MM$AS. | 3.5 |

### 4.3 Empirical Methodology

Eleven software development professionals with experience of early lifecycle software design were invited to participate in trials using the proposed iACO approach. Relevant information concerning their background is given in Appendix 1. The total experience of software development of the participants amounts to 228 years in both academia and industrial practice. Participants 4 and 9 are authors of this paper. Details of the Research Ethics process can be found at [55]. In brief, the iACO approach is explained to participants and use of the iACO environment is presented using a dummy design problem. Each of the three software design problems is then described. Once underway, each interactive design episode is allowed to proceed until the participant decides to halt. However, to prevent user fatigue, each participant session is curtailed after one hour whether or not the planned schedule of five episodes had been completed.

In order to test the effect of design problem (CBS, GDP and SC), the 'freeze' and 'archive' capability, as well as the effect of the color scheme, an experimental schedule of five episodes was devised and is shown in Appendix 2. At each ACO iteration, a record is stored containing enough details to fully identify the specific run, along with the best values for CBO, NAC and ATMR achieved by the colony in that iteration. In addition, at each ACO iteration where designer interaction occurs, all details of the user's interaction (value of evaluation, classes frozen/unfrozen, archive) along with the updated values for the weights of CBO, NAC and ATMR are also recorded. Lastly, at the end of the iACO design session, each participant is invited to provide any comments on their overall human experience of the trial. Such comments might include any satisfying aspects, any aspects that generated user fatigue, and any suggestions for enhancement of the overall human experience.

## 5 RESULTS

All experimental data are available at [56]; we next report key findings.

## 5.1 Number of Interactions

Table 3 shows the number of interactions during design episodes for each participant and each design problem. Where a participant did not conduct a design episode due to time constraints, this is shown as "-". Participants evaluated candidate software designs on a total of 962 interactions. Immediately apparent is the great variation in the number of interactions among the participants, reflecting a variety of individual approaches. Numbers for CBS and GDP are higher than SC as the experimental design meant that most participants undertook two design episodes for these design problems. Thus to analyze these figures, the numbers of interactions for each design problem have been examined further, and the results are summarized in Table 4 where standard deviations are shown in parentheses. The highlights of Table 4 are twofold: firstly, there is a high variation in number of interactions for the CBS and GDP design problems when compared to SC, and secondly, the mean number of interactions for CBS and GDP are similar and much higher than that for SC. Wilcoxon Signed Ranks Test confirms that while the differences between CBS and SC, and GDP and SC are significant ($p = .027$ and $p = .028$ respectively), the differences between CBS and GDP are not. To explain these outcomes, if we look to the numbers of classes in each of the design problems, we find that the number of classes in candidate design solutions for CBS and GDP is 5, whereas for SC the number is 15. Therefore, it seems likely that the cognitive load on the software design is higher for the SC design problem, accounting for the significant differences in the number of interactions.

## 5.2 Example Fitness Values

A typical example of the fitness values curves achieved in an interactive iACO design episode is shown in Figure 2. A mid-scale design problem i.e. GDP has been chosen for illustration from a design episode for Participant 2. Figure 2 shows that the iACO search engine appears highly effective in achieving superior fitness values for all three design measures, performing well within 35 iterations. Thereafter, at the end of the design episode, all three fitness measures are superior to values for the corresponding manual design. However, while this is a typical example, a degree on variation in the design episodes has also been observed, not least in the number of iterations reached before halting.

Table 3. Number of Interactions for each Participant.

|             | Design Problem |     |    |       |
| Participant | CBS | GDP | SC | Total |
|---|---|---|---|---|
| 1  | 98  | 149 | 12 | 259 |
| 2  | 36  | 30  | -  | 66  |
| 3  | 47  | 29  | 13 | 89  |
| 4  | 35  | 13  | 8  | 56  |
| 5  | 44  | 107 | 17 | 168 |
| 6  | 36  | 18  | 10 | 64  |
| 7  | 45  | -   | -  | 45  |
| 8  | 17  | 6   | -  | 23  |
| 9  | 27  | 27  | -  | 54  |
| 10 | 30  | 32  | 12 | 74  |
| 11 | 64  | -   | -  | 64  |
| Total | 479 | 411 | 72 | 962 |

Table 4. Numbers of Interactions for each Design Problem.

|     | N  | Minimum | Maximum | Mean |
|-----|----|---------|---------|------|
| CBS | 11 | 17 | 98  | 43.545 (21.786) |
| GDP | 9  | 6  | 149 | 45.666 (48.610) |
| SC  | 6  | 8  | 17  | 12.000 (3.033)  |

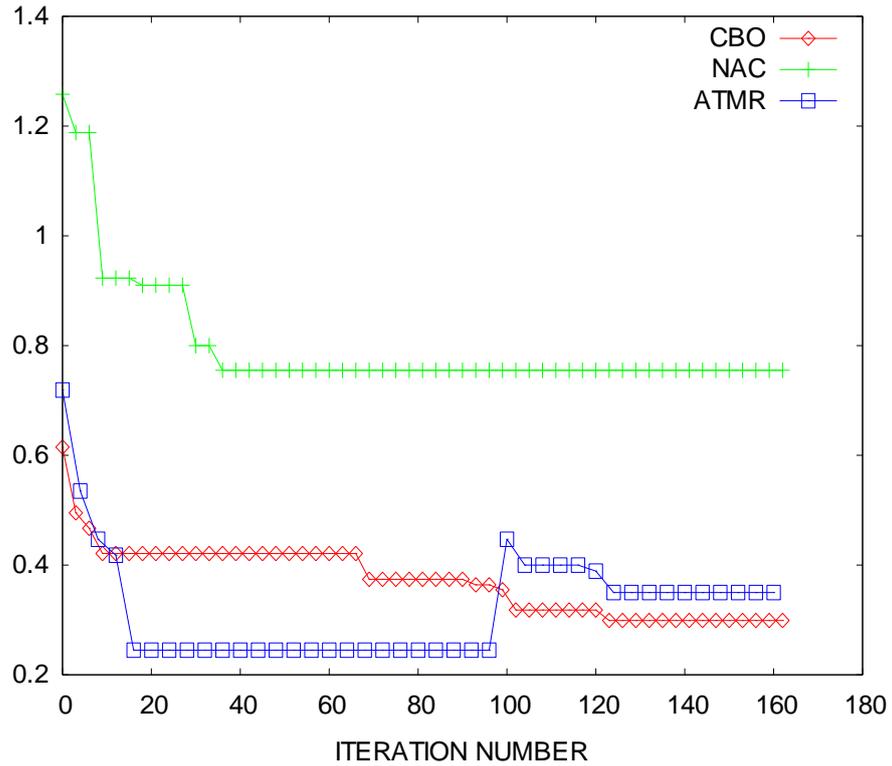

Figure 2. Progression of Fitness Values for CBO, NAC and ATMR in an iACOdesign episode for a mid-scale design problem (GDP).

### 5.3 Variation in Fitness Values at End of Episodes

Table 5 shows the best values obtained for the three fitness metrics at the last interaction of each participant episode. In table 5, 'N' indicates the number of participant episodes at the end of which fitness values have been recorded. The 'Best' row shows the single best value achieved in all episodes for each design problem, while the 'Mean' row shows the mean of all best values at the end of episodes for each design problem, with the standard deviation shown in parentheses. Fitness values for the manually produced software designs are shown in italic font for comparison. Bold font is used to indicate that fitness values achieved (either single best or mean best) using iACO are superior to those of the manually produced design. To establish if the differences between mean values are statistically significant, the single sample t-test has been used to compare the sample means (i.e. of the manual designs) against the means for the target designs.

Table 5. Best fitness values for CBO, NAC, ATMR at end of participant episodes.

| | | Design Problem | | | | | |
|---|---|---|---|---|---|---|---|
| | | CBS (N=22) | | GDP (N=17) | | SC (N=6) | |
| CBO | Best | **0.175** | | **0.234** | | 0.562 | |
| | *Manual* | *0.154* | | *0.297* | | *0.452* | |
| | Mean | 0.265 | (0.045) | 0.298 | (0.062) | 0.602 | (0.029) |
| | T-Test | $p < .001$ | | | | $p < .001$ | |
| NAC | Best | **0.200** | | **0.490** | | **1.038** | |
| | *Manual* | *0.821* | | *2.592* | | *1.520* | |
| | Mean | 1.599 | (1.291) | **1.902** | (1.966) | **1.292** | (0.169) |
| | T-Test | | | | | $p = .222$ | |
| ATMR | Best | **0.036** | | **0.249** | | **0.406** | |
| | *Manual* | *0.199* | | *2.617* | | *1.848* | |
| | Mean | 0.044 | (0.040) | 0.679 | (0.333) | 0.602 | (0.110) |
| | T-Test | $p < .001$ | | $p < .001$ | | $p < .001$ | |

For the sake of brevity, *p* values are only shown where differences are significant at the alpha = 0.05 level. Analyzing each fitness measure in turn, we firstly see that for CBO, mean values for CBS and SC are a little inferior to values for the manually produced design, and this difference is statistically significant. However, the mean CBO value for the GDP problem is very similar to that produced for the manual design, and the best CBO value is superior. Secondly, for NAC, it is evident that the best value achieved is superior to the manual design value for all design problems, and the mean values are also superior for GDP and SC, the difference being statistically significant for the SC problem. Thirdly, for the ATMR metric, all best values and the mean values are superior for all design problems, and the differences for the mean values are statistically significant.

Overall, the results appear to indicate that candidate design solutions produced by participants using the iACO environment can be superior to the manually produced design with respect to NAC and ATMR values, although a little inferior for CBO. With regard to design problem, results obtained for GDP are excellent, but although still good, perhaps less so for CBS and SC. The character of the results may be to some extent explained by the multi-objective nature of the design evaluation, and the increased scale of the SC problem (15 classes to the 5 for CBS and GDP). Overall, these results are interesting, and appear to suggest not only that iACO is effective overall in searching for software design solutions, but also that elegance does indeed play an important role in software design. Of course, the superior elegance values arrived at during participant designer episodes could be caused by a number of factors, not least the multi-objective value judgments made by the participants. However, it is also highly likely that these results are influenced by the iACO learning mechanisms during interactive search, and this is discussed in section 5.5.

## 5.4 Effect of Designer Hints

To examine the effect of freezing and color scheme, we conducted a 2 x 2 mixed analysis of variance with freezing (on, off) as a 2 level between subjects variable and color scheme (traffic lights, water tap) as a 2 level repeated measures factor with outcomes CBO and NAC at the last designer interaction. However, there are two important considerations in our analysis. Firstly, because the sample size is restricted, the largest design problem CBS (N=22) has been chosen as for analysis. GDP and SC, with sample sizes of 17 and 6 respectively, have therefore not been analyzed. (Of course, if significant results are not obtained for CBS, there seems little point in proceeding to analyze GDP and SC.) Secondly, we find that the ATMR data presents a curious distribution. Of the 22 data values for ATMR at the end of iACO design episodes for the CBS, further inspection reveals that the value 0.036 presents 20 times. Indeed, there are only 2 discrepant values, i.e. 0.224 and 0.044 (which explains the low standard deviation obtained for ATMR in table 6). This suggests that ATMR is less sensitive as a measure in the multi-objective evaluation performed by participants in this investigation, and possible causes and consequences of this are discussed in the following sections.

For both CBO and NAC, the analysis reveals no statistically significant differences between results obtained with freezing on and freezing off, or for the color scheme used. Nevertheless, it does appear that when freezing is on, better results are obtained with the water tap color scheme. On the other hand, when freezing is off, it appears that better results are obtained with the traffic light color scheme. However, mixed analysis of variance indicates that this potential statistical interaction effect is not statistically significant. Thus while this appearance is indicative of the effect of freezing and color scheme, it is not conclusive. In an attempt to explain these findings, we suggest that the variability in participant interaction with the iACO environment for the given sample size is a factor. It was also observed that while some participants made heavy use of the freeze capability, others did not despite being aware of its presence. With regard to color scheme, participants seemed able to use both effectively, and results of the participant questionnaire are reported in section 5.6, "Human Experience".

## 5.5   Learning of Fitness Weights

Mean values of the weights for CBO, NAC and ATMR ($W_{CBO}$, $W_{NAC}$, $W_{ATMR}$ respectively) learned by the iACO environment at the final interaction at the end of episodes are shown with standard deviation in parentheses in table 6. Table 6 reveals the overall balance obtained between the learned weights, and also the impact of scale of design problem. Firstly, it is evident that $W_{CBO}$ emerges as the highest learned weight for all three design problems. It is also evident that $W_{NAC}$ appears as the lowest learned weight overall, although not for SC. This strongly suggests that the balance between the three learned weights is problem dependent.

Table 6. Weight Values for CBO, NAC, ATMR at end of Participant Episodes.

| Design Problem | $W_{CBO}$ | $W_{NAC}$ | $W_{ATMR}$ |
| --- | --- | --- | --- |
| CBS (N=22) | 0.588  (0.208) | 0.097  (0.058) | 0.314  (0.233) |
| GDP (N=17) | 0.742  (0.251) | 0.075  (0.062) | 0.182  (0.227) |
| SC (N=6) | 0.817  (0.073) | 0.096  (0.073) | 0.086  (0.063) |
| Total (N=45) | 0.677  (0.229) | 0.088  (0.061) | 0.233  (0.233) |

Secondly, we see that $W_{NAC}$ is similar across all scales of design problem whereas $W_{CBO}$ increases and $W_{ATMR}$ decreases with scale. We conjecture that as the cognitive load of the design problem increases, the iACO environment learns that participants are placing less emphasis on qualitative design elegance and rely more on the quantitative measure of Coupling Between Objects (CBO).

To further explain the above characteristics, we draw on the findings related to the ATMR measure discussed in previous sections, wherein the data suggested that ATMR is not as sensitive a measure at CBO or NAC. Interestingly, it was observed during design episodes that from time-to-time, the participants were presented with visualizations of candidate software design solution paths that show the "God Class" anti-pattern [57]. This is generally regarded by software designers as a most inelegant design solution to be avoided, wherein a single class acts as an incoherent grouping of a large number of attributes and methods, typically leaving other classes with ineffectively small numbers. An example of a "God Class" solution for the GDP problem is shown in Figure 3. The values of CBO, NAC and ATMR fitness for this solution are 0.439 (0.297), 8.691 (2.592) and 0.194 (0.199) respectively, with values from the manual design in parentheses for comparison. It is evident that for this candidate software design, although CBO and ATMR are approaching or better than the manual design, the value of NAC is inferior. It is, of course, fundamental to multi-objective search that the fitness metrics used in search conflict. However, when a God Class is present, it seems likely that the CBO and ATMR metrics are not in conflict, as the God Class creates high values for both measures. To help shed further light on this behavior, Figure 4 shows the best fitness value curves obtained when a 'God class' comes to predominate a design solution. This shows that best fitness values for all three measures initially progress well until generation 95 when a simultaneous improvement in CBO and dramatic decrease in NAC are evident. At this point, it seems possible that the improvement in CBO comes at the expense of NAC, although ATMR appears steady. This behavior appears consistent with (i) ATMR being less sensitive and (ii) a lack of sufficient conflict between the CBO and ATMR measures.

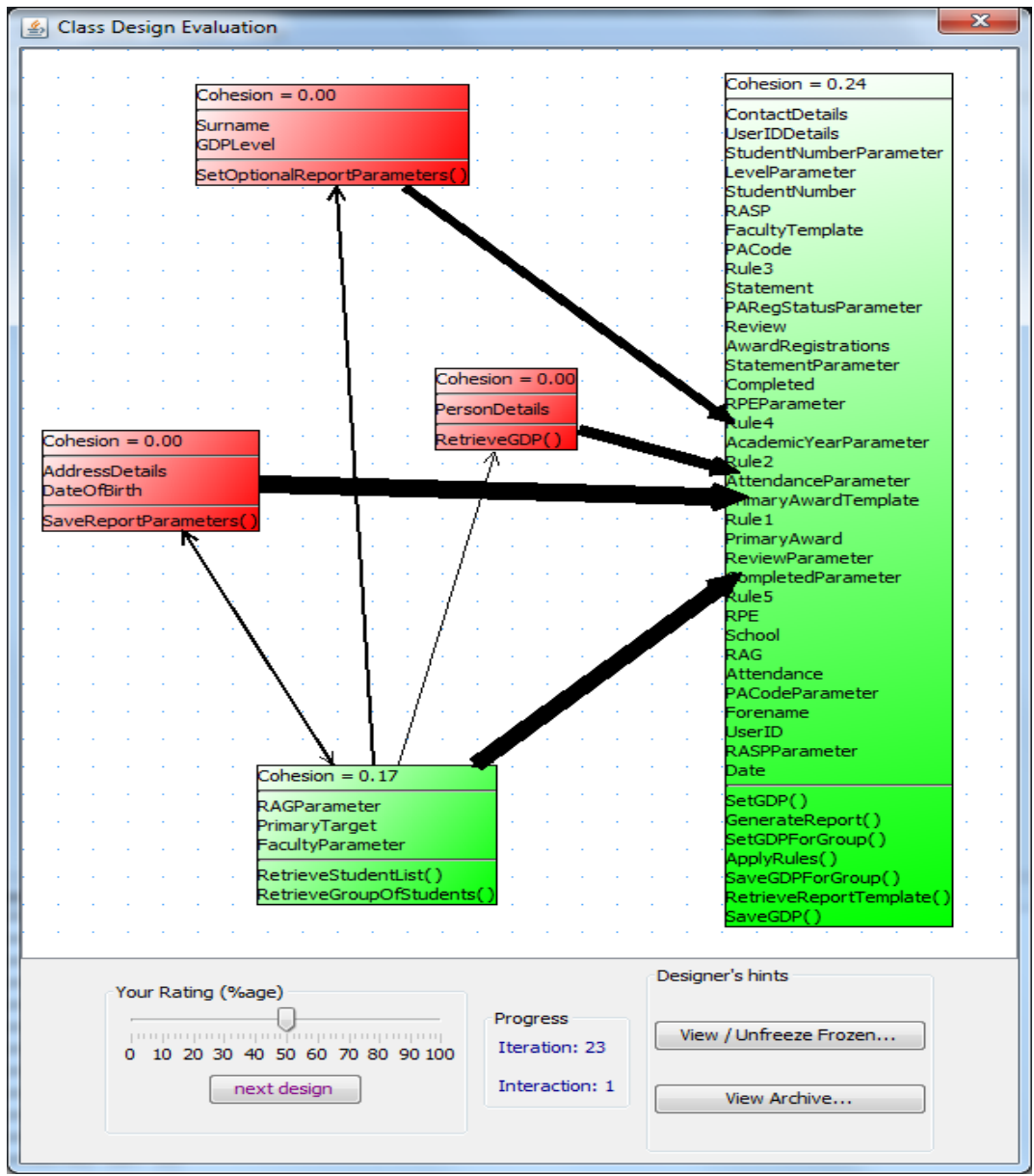

Figure 3. Example of candidate software design solution with a "God Class".

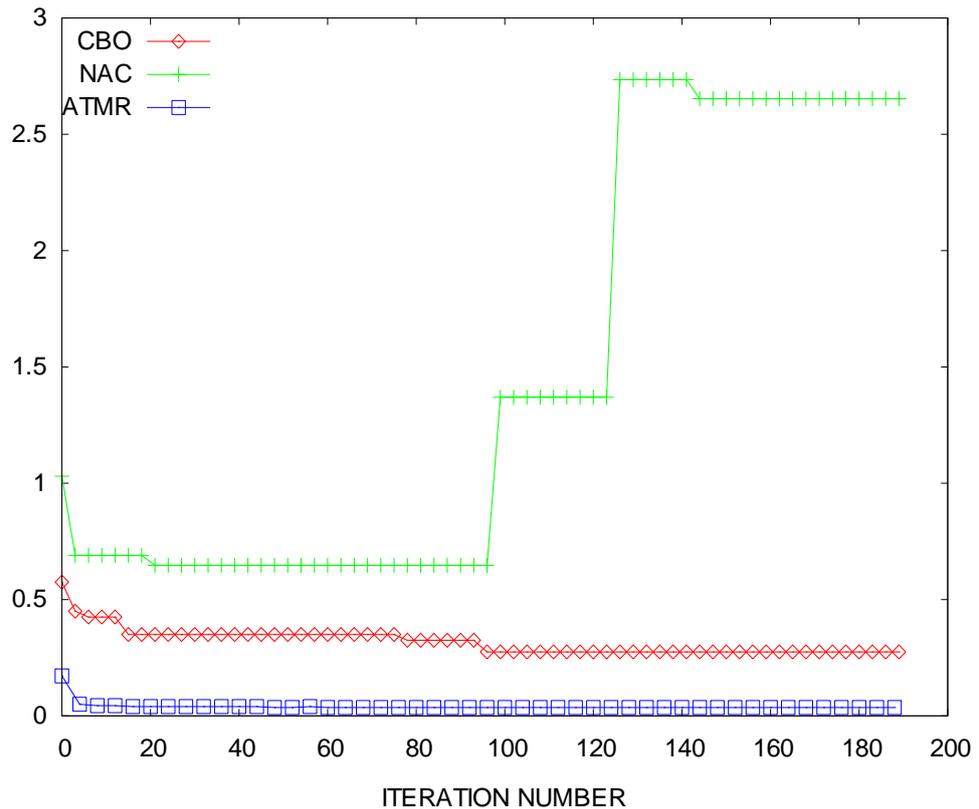

Figure 4. Example of best fitness value curves for a software design solution with a "God Class".

**Together, the results of Figures 3 and 4 suggest that the influence of NAC on steering the search can be diminished, which accounts for the low $W_{NAC}$ values learned by the iACO environment. Furthermore, given the lack of sensitivity of ATMR as a surrogate elegance**

**measure, we conjecture that in future work, a bi-objective search using CBO and NAC measures alone can be effective to achieve elegant candidate software design solutions with interactive ACO. 5.6 Human Experience**

Ten of the eleven participants responded to the questionnaire [55] and the results are as follows. Asked to rate how compelling they found the interactive ACO design experience on a scale from 1 ("Not at all compelling") to 5 ("Very Compelling"), five participants rated the interactive design experience at 5, while the other five participants rated the experience at 4. We applied 95% confidence levels for proportion (using the Pearson Clopper intervals) and found this to be a statistically significant positive rating ($p = .002$). Participants were also asked to rate how effective they found the iACO design experience at achieving useful and relevant software design solution paths, (scale as before with "effective" replacing "compelling") Three participants rated the effectiveness at 5, four participants rated the effectiveness at 4, and a further three participants rated the effectiveness at 3. Although 7 ratings are positive and three ratings are neutral, 95% confidence levels for proportions does not show statistical significance. We conjecture that this is consistent with, and reflects the participants' perception of the findings in the previous section. It seems possible that although the iACO environment achieves design solutions of superior fitness, the lack of sensitivity of the ATMR metrics might be implicitly perceived as constraining the effectiveness of interactive search.

When asked to comment on their preferred color scheme, 7 out of 10 participants stated a preference for 'traffic lights', whereas the remaining 3 participants expressed a preference for the 'water tap' scheme. Although this indicates a greater preference for the 'traffic lights' scheme, the results of the previous section seem to indicate that the color scheme does not have a statistically significant effect on the participant performance using iACO environment. We also found no statistically significant difference between the mean values for CBO, NAC and ATMR between the preferred and the least preferred color scheme. This suggests that participants perform well with either color scheme and that the iACO search is robust with respect to the implementation of color scheme visualization.

Many of the "free text" participant comments about the iACO experience were positive e.g. *"the tool looks good and works well"*, and *"the tool did seem to help quickly arrive at an optimal class design"*. Other participants commented on the effectiveness of the design visualization e.g. *"the visibility of the cohesion and coupling"* and the use of a color scheme that *"speeded up the decision process"*. When asked for suggestions for improving the iACO experience, participants suggested even more interactivity, such as a visual indication of a frozen class (perhaps an ice cube icon on top of the class), the ability to backtrack along the history of the episode and restarting search from a particular design variant, and the capability to hint to the iACO environment by 'drag and drop' of attributes and methods from one class grouping to another.

## 6      THREATS TO VALIDITY

With respect to the interval validity of results, the iACO design experience is highly dependent on the design context, and so every attempt has been made to make a consistent design context for all participants. The same briefing has been received by all participants and all trials have been conducted in the same iACO environment. An additional threat to internal validity is the Hawthorn affect, in which participant behavior may be changed by the special situation and social treatment they received during the experiment. To counter this, participation was conducted as consistently as possible; furthermore it was explained to participants that the halting of interactive design episodes was entirely at their discretion and that there was no expectation in relation to the particular designs arrived at.

Two other threats to internal validity include the learning affect and the fatigue affect. The learning effect threatens validity in the sense that participant capability improves during the episodes through learning by repetition. To counter this, the experimental setup includes a period of familiarization with a dummy design problem first, so that knowledge of how to use the iACO environment is instilled prior to proceeding with the three design problems. The fatigue effect is mitigated by ensuring that design episodes are halted after one hour's duration.

With respect to external validity of results, the outcomes of the investigations depend on the number and experience of the participants being representative of some segment of the software design community. The 228 years' experience of professional software development among the 11 participants includes 149 years of academic experience. It also includes 79 years of industrial software design and development experience for participants 1, 7 and 10 who have architected and developed software across a wide variety of software design domains, within object-oriented and service-based technical architectures worldwide. While a greater number of participants would have lent greater robustness to the statistical analysis of the study, the years of experience of the trial participants suggests a level of credibility for their evaluations of the candidate software designs presented by the interactive ACO environment.

# 7　CONCLUSIONS

As judged from the quantitative results and participant feedback, we conclude that ACO is highly effective as a search engine for interactive, dynamic multi-objective interactive search in early lifecycle software design. Indeed, with speedy discovery of useful candidate software design solution paths, study participants rate the interactive ACO search experience as compelling. While the results into the influence of color scheme and designer 'hints' such a freezing have proved statistically inconclusive, the sample size is relatively small and great variation in participant behavior during interaction is evident. Nevertheless, study participants have provided positive ratings and comments for both 'hint' capabilities, and we plan their incorporation in any future investigations.

Results of machine learning of fitness measure weightings are interesting and indicate that software design elegance does indeed play a significant role in designer evaluation of candidate software design. Furthermore, we conclude that the surrogate elegance measure of the ratio of attributes to methods (ATMR) is less effective in multi-objective search, as it fails to steer the search away from the "God class" anti-pattern. This is significant as it seems likely that the evenness of the distribution of attributes and methods among classes (NAC) is the more significant surrogate elegance measure, which in turn suggests that this evenness of distribution, when combined with structural integrity, is an implicit but crucial component of effective early lifecycle software design.

# REFERENCES

[1] Simons, C.L., Parmee, I.C. (2009) An Empirical Investigation of Search-based Computational Support for Conceptual Software Engineering Design. In: *Proceedings of the 2009 IEEE International Conference on Systems, Man, and Cybernetics, (SMC '09),* pp. 2577-2582.

[2] Simons, C.L., Parmee, I.C., Gwynllyw, R. (2010) Interactive, Evolutionary Search in Upstream Object-oriented Software Design. *IEEE Transactions on Software Engineering,* vol. 33, no. 6, pp. 798-816.

[3] Simons, C.L., Parmee, I.C. (2012) Elegant, Object-Oriented Software Design via Interactive Evolutionary Computation. *IEEE Trans. Systems, Man, and Cybernetics: Part C – Applications and Reviews.* In Press.

[4] Simons, C.L., Smith, J.E. (2012) A Comparison of Evolutionary Algorithms and Ant Colony Optimisation for Interactive Software Design. In: *Fast Abstract Collection of the 4th Symposium of Search-Based Software Engineering, (SSBSE 2012),* pp. 37-42.

[5] Simons, C.L., Smith, J.E. (2012) A Comparison of Meta-heuristic Search for Interactive Software Design. *Soft Computing*. Submitted. Pre-publication version available at http://arxiv.org/abs/1211.3371

[6] Xanthakis, S., *et al.*, (1992) Application of Genetic Algorithms to Software Testing. In: *Proceedings of the 5th International Conference on Software Engineering,* pp. 625-636.

[7] Smith, J.E., Fogarty, T.C. (1996) Evolving Software Test Data - GAs Learn Self-expression. *Evolutionary Computing*, ed. Fogarty, T.C., Springer, pp. 137-146.

[8] Harman, M., Jones, B.J. (2001) Search-Based Software Engineering. *Information and Software Technology,* vol. 43, no.14, pp. 833-839.

[9] Ren, J., Harman, M., Di Penta, M. (2011) Cooperative Co-evolutionary Optimisation of Software Project Assignments and Job Scheduling. In: *Proceedings of the 3rd International Symposium of Search Based Software Engineering (SSBSE 2011), LNCS 6956*, pp. 127-141.

[10] Bowman, M., Briand, L.C., Labiche, Y. (2010) Solving the Class Responsibility Assignment Problem in Object-Oriented Analysis with Multi-objective Genetic Algorithms. *IEEE Transactions in Software Engineering,* vol. 36, no. 6, pp. 817–837.

[11] McMinn, P. (2004) Search-Based Software Test Data Generation: a Survey. *Software Testing, Verification and Reliability*, vol. 14, no 2, pp. 105-156.

[12] Weimer, W., Forrest, S., Le Goues, C., Nguyen, T. (2010) Automatic Program Repair with Evolutionary Computing. *Communications of the ACM,* vol. 53, no. 5, pp. 109-116.

[13] O'Keeffe, M., Cinneide, M.O. (2008) Search-Based Refactoring for Software Maintenance. *Journal of Systems and Software*, vol. 81, no. 4, pp. 502-516.

[14] Zhang, Y. (2012) Repository of Publications on Search-based Software Engineering. http://crestweb.cs.ucl.ac.uk/resources/sbse_repository/ Accessed 23 October 2012.

[15] Cockburn, A. (2002) *Agile Software Development.* Addison-Wesley.

[16] Martin, R.C. (2003) *Agile Software Development: Principles, Patterns and Practices.* Prentice-Hall.

[17] Maiden, N. (2011) Requirements and Aesthetics. *IEEE Software*, vol. 28, no. 3, pp. 20-21.

[18] Object Management Group. *Unified Modelling Language Resource Page*. http://www.uml.org/ Accessed 23 October 2012.

[19] Takagi, H. (2001) Interactive Evolutionary Computation: Fusion of the Capabilities of EC Optimization and Human Evaluation. *Proceedings of the IEEE*, vol. 89, no. 9, pp. 1275–1298.

[20] Caldwell, C., Johnston, V.S. (1991) Tracking a Criminal Suspect through "Face-Space" with a Genetic Algorithm. In: *Proceedings of the 4th International Conference on Genetic Algorithms*, pp. 416-421.

[21] Sims, K. (1991) Artificial Evolution for Computer Graphics. *Computer Graphics, vol. 25, no. 4, SigGraph '91 Proceedings*, pp. 319-328.

[22] Legrand, P., Bourgeois-Republique, C., Pean, V., Harboun-Cohen, E., Levy-Vehel, J., Frachet, B., Lutton, E., Collet, P. (2007) Interactive Evolution for Cochlear Implants Fitting. *Genetic Programming and Evolvable Machines*, vol. 8, no. 4, pp. 301-318.

[23] Miettinen, K.M. (1998) *Nonlinear Multiobjective Optimization*. Kluwer.

[24] Belton, V., Branke, J., Eskelinen, P., Greco, S., Molina, J., Ruiz, F., Slowinski, R. (2008) Interactive Multiobjective Optimization from a Learning Perspective. In: *Multiobjective Optimization: Interactive and Evolutionary Approaches, LNCS 5252,* pp. 405-433.

[25] Deb, K. (2012) Advances in Evolutionary Multi-Objective Optimization. In: *Proceedings of the 4th International Symposium on Search-Based Software Engineering (SSBSE), LNCS 7515,* pp. 1-26.

[26] Caleb-Solly,P., Smith, J.E. (2007) Adaptive Surface Inspection via Interactive Evolution, *Image and Vision Computing,* vol. 25, no. 7, pp. 1058-1072.

[27] Ohsaki, M., Takagi, H., Ohya, K. (1998) An Input Method using Discrete Fitness Values for Interactive Genetic Algorithms. *Journal of Intelligent and Fuzzy Systems*, vol. 6, no. 1, pp. 131-145.

[28] Miller, G. (1956) The Magical Number Seven, Plus or Minus Two: Some Limits on our Capacity for Processing Information. *Psychology Review,* vol. 63, no. 2, pp. 81-97.

[29] Lee, J.-Y., Cho, S.-B. (1999) Interactive Genetic Algorithm with Wavelet Coefficients for Emotional Image Retrieval. In: *Proceedings of the 5th International Conference on Soft Computing and Information /Intelligent Systems*, volume 2, pp. 829-832.

[30] Boudjeloud, L., Poulet, F. (2005) Visual Interactive Evolutionary Algorithm for High Dimensional Data Clustering and Outlier Detection. In: *Proceedings of the 9th Pacific-Asia Conference on Advances in Knowledge Discovery and Design,* pp. 428-431.

[31] Kubota, N., Nojima, Y., Kojima, F., Fukuda, T. (2006) Multiple Fuzzy State-value Functions for Human Evaluation through Interactive Trajectory Planning of a Partner Robot. *Soft Computing,* vol. 10, no. 10, pp. 891-901.

[32] Avigad, G., Moshaiov, A., Brauner, N. (2005) Interactive Concept-based Search using MOEA: the Hierarchical Preference Case. *International Journal of Computational Intelligence*, vol. 2, no. 3, pp. 182-191.

[33] Brintrup, A., Ramsden, J., Takagi, H., Tiwari, A. (2008) Ergonomic Chair Design by Fusing Qualitative and Quantitative Criteria using Interactive Genetic Algorithms. *IEEE Transactions on Evolutionary Computation*, vol. 12, no. 3, pp. 343–354.

[34] Eiben, A.E., Smith, J.E. (2003) *Introduction to Evolutionary Computing*, Springer.

[35] Dorigo, M., Di Caro, G. (1999) Ant Algorithms for Discrete Optimization. *Artificial Life*, vol. 5, no. 2, pp. 137-172.

[36] Dorigo, M., Stutzle, T. (2004) *Ant Colony Optimization.* MIT Press.

[37] Bilchev, G., Parmee, I.C. (1995) The Ant Colony Metaphor for Search Continuous Design Spaces. *Evolutionary Computing: Lecture Notes in Computer Science 993,* pp. 25-39.

[38] Madar, J., Abonyi, J., Szeifert, F. (2005). Interactive Particle Swarm Optimisation. In: *Proceedings of the 5th International Conference on Intelligent Systems Design and Applications (ISDA '05),* pp. 314-319.

[39] Xing, L.-N., Chen,Y.-W., Yang, K.-W. (2007) Interactive Fuzzy Multi-Objective Ant Colony Optimisation with Linguistically Quantified Decision Functions for Flexible Job Shop Scheduling Problems. In: *Proceedings of Frontiers in the Convergence of Bioscience and Information (FBIT 2007),* pp. 801-806.

[40] Uğur, A., Aydin, D. (2009) An Interactive Simulation and Analysis Software for Solving TSP using Ant Colony Optimization Algorithms. *Advances in Engineering Software*, vol. 40, no. 5, pp. 341–349.

[41] Albakour, M.-D., Kruschwitz, U., Nanas, N., Song, D., Fasli, M., De Roeck, A. (2011) Exploring Ant Colony Optimisation for Adaptive Interactive Search. In: *Proceedings of Advances in Information Retrieval Theory, LNCS 6931*, pp. 213-224.

[42] Toth, P., Vigo, D. (2001) *The Vehicle Routing Problem.* Philadelphia.

[43] Harrison, R., Councell, S., Nithi, R. (1998) An Investigation into the Applicability and Validity of Object-oriented Design Metrics. *Empirical Software Engineering,* vol. 3, no. 3, pp. 255-273.

[44] Alaya, I., Solnon, C. Ghedira, K. (2007) Ant Colony for Multi-Objective Optimisation Problems. In: *Proceedings of the 19th IEEE International Conference on Tools with Artificial Intelligence,* pp.450-457.

[45] Dorigo, M., Stutzle, T. (2010) Ant Colony Optimisation: Overview and Recent Advances. *Handbook of Metaheuristics: International Series in Operations Research and Management Science,* vol. 146, pp.227-263, Springer.

[46] Lopez-Ibanez, M., Stutzle, T. (2012) An Experimental Analysis of Design Choices for Multi-Objectives Ant Colony Optimisation Algorithms. *Swarm Intelligence,* vol. 6, no. 3, pp. 207-232.

[47] Stutzle, T., Hoos, H. (2000) MAX-MIN Ant System. *Future Generation Computer Systems,* vol. 16, no. 8, pp. 889-914.

[48] Simons, C.L. (2012) Example Software Design Visualizations, http://www.cems.uwe.ac.uk/~clsimons/iACO/ExampleVisualisations.pdf, Accessed 12 November 2012.

[49] Buchanan, J.T., Daellenbach, H.G. (1997) The Effects of Anchoring in Interactive MCDM Solution Methods. *Computers and Operations Research,* vol. 24, no. 10, pp. 907-918.

[50] Eiben, A.E., Smith, J.E. (2003) *An Introduction to Evolutionary Computing,* Springer.

[51] Simons, C.L. *Use Case Specifications for Cinema Booking System*, http://www.cems.uwe.ac.uk/~clsimons/CaseStudies/CinemaBookingSystem.htm Accessed 31 October 2012.

[52] Simons, C.L. *Use Case Specifications for Graduate Development Program*, http://www.cems.uwe.ac.uk/~clsimons/CaseStudies/GraduateDevelopmentProgram.htm Accessed 31 October 2012.

[53] Simons, C.L. *Use Case Specifications for Select Cruises*, http://www.cems.uwe.ac.uk/~clsimons/CaseStudies/SelectCruises.htm Accessed 31 October 2012.

[54] Simons, C.L. *Manual Software Designs for Problem Domains,* http://www.cems.uwe.ac.uk/~clsimons/CaseStudies/ManualDesigns.pdf Accessed 31 October 2012.

[55] Simons, C.L. *Ethical Approval Documents*, http://www.cems.uwe.ac.uk/~clsimons/Ethics/ Acessed 31 October 2012.

[56] Simons, C.L. Raw Experimental Data, http://www.cems.uwe.ac.uk/~clsimons/iACO/ Accessed 6 November 2012.

[57] Brown, W.J., Malveau, R.C., McCormick III, H.W., Mowbray, T.J. (1998) *Anti-Patterns: Refactoring Software, Architectures, and Projects in Crisis,* Wiley.

# Appendix 1: Participants

Details of the gender, profession and years' experience of each participant of the study are as follows:

| Participant | Gender | Current Profession | Years in Industry | Years in Academia | Total Years |
|---|---|---|---|---|---|
| 1 | Male | Lecturer | 31 | 2 | 33 |
| 2 | Male | Lecturer | 12 | 23 | 35 |
| 3 | Female | Lecturer | 5 | 24 | 29 |
| 4 | Male | Researcher | 1 | 20 | 21 |
| 5 | Male | Researcher | 2 | 16 | 18 |
| 6 | Male | Designer | 0 | 20 | 20 |
| 7 | Male | Engineer | 6 | 19 | 25 |
| 8 | Male | Help Desk | 1 | 0 | 1 |
| 9 | Female | Lecturer | 10 | 13 | 23 |
| 10 | Male | Lecturer | 10 | 12 | 22 |
| 11 | Male | Student | 1 | 0 | 1 |
|  |  | TOTAL | 79 | 149 | 228 |

# Appendix 2: Participant Schedule

In the following participant schedule, TL refers to the 'Traffic Lights' color scheme, while WT refers to the 'Water Tap' color scheme.

| Participant | Problem | Episode 1 CBS | 2 CBS | 3 GDP | 4 GDP | 5 SC |
|---|---|---|---|---|---|---|
| 1 | Freeze | off | off | off | on | on |
|   | Color | TL | WT | any | any | TL |
| 2 | Freeze | off | off | on | off | off |
|   | Color | WT | TL | any | any | TL |
| 3 | Freeze | off | off | off | on | On |
|   | Color | TL | WT | any | any | WT |
| 4 | Freeze | off | off | on | off | off |
|   | Color | WT | TL | any | any | WT |
| 5 | Freeze | off | off | off | on | on |
|   | Color | TL | WT | any | any | TL |
| 6 | Freeze | on | on | off | on | On |
|   | Color | TL | WT | any | any | WT |
| 7 | Freeze | on | on | on | off | off |
|   | Color | WT | TL | any | any | WT |
| 8 | Freeze | on | on | off | on | on |
|   | Color | TL | WT | any | any | TL |
| 9 | Freeze | on | on | on | off | off |
|   | Color | TL | WT | any | any | TL |
| 10 | Freeze | on | on | off | on | on |
|   | Color | TK | WT | any | any | WT |
| 11 | Freeze | on | on | on | off | off |
|   | Color | WT | TL | any | any | WT |